\documentclass[conference]{IEEEtran}  
\usepackage{ifpdf}
\usepackage{cite}
\usepackage{graphicx}
\usepackage{epstopdf}
\usepackage{amssymb}

\ifCLASSINFOpdf
\else
\fi

\usepackage[cmex10]{amsmath}

\usepackage{color}

\begin{document}
\title{An Exchange Mechanism to Coordinate Flexibility in Residential Energy Cooperatives}


\author{\IEEEauthorblockN{Shantanu Chakraborty}
\IEEEauthorblockA{The University of Melbourne, Melbourne, Australia.\\
shantanu.chakraborty@unimelb.edu.au}
\and
\IEEEauthorblockN{Pablo Hernandez-Leal and Michael Kaisers}
\IEEEauthorblockA{Centrum Wiskunde \& Informatica, Amsterdam, The Netherlands.\\
\{Pablo.Hernandez,Michael.Kaisers\}@cwi.nl}
}
\maketitle

\begin{abstract}
Energy cooperatives (ECs) such as residential and industrial microgrids have the potential to mitigate increasing fluctuations in renewable electricity generation, but only if their joint response is coordinated. 
However, the \emph{coordination} and \emph{control} of independently operated flexible resources (e.g., storage, demand response) imposes critical challenges arising from the heterogeneity of the resources, conflict of interests, and impact on the grid. 
Correspondingly, overcoming these challenges with a general and fair yet efficient exchange mechanism that coordinates these distributed resources will accommodate renewable fluctuations on a local level, thereby supporting the energy transition.
In this paper, we introduce such an exchange mechanism. It incorporates a payment structure that encourages prosumers to participate in the exchange by increasing their \emph{utility} above baseline alternatives. The allocation from the proposed mechanism increases the system efficiency (\emph{utilitarian} social welfare) and distributes profits more fairly (measured by \emph{Nash} social welfare) than individual flexibility activation.
A case study analyzing the mechanism performance and resulting payments in numerical experiments over real demand and generation profiles of the Pecan Street dataset elucidates the efficacy to promote cooperation between co-located flexibilities in residential cooperatives through local exchange.

\end{abstract}

\begin{IEEEkeywords}
Exchange Mechanism, Energy Cooperative, Multi-agent System, Flexibility, Market Design.
\end{IEEEkeywords}


\section{Introduction}



Energy Cooperatives (ECs) of prosumers are gaining considerable traction due to their potentials for efficient energy management, reduced grid dependency, and increased local consumption of distributively generated renewable energy.
Establishing a local market can facilitate the management of distributed renewable generation that needs to consume as \emph{locally} as possible~\cite{rosen2014regulatory}. 

A generic \emph{prosumer marketplace} can be envisaged as a multiagent system that hosts heterogeneous services with a large number of participant prosumers which take actions autonomously~\cite{rodriguez2014business}. This autonomy and diversity form a complex system where efficient mechanisms such as local markets are crucial~\cite{Lezama:2018}. The necessity of establishing a local \emph{prosumer marketplace} (in the sense of the scope and proximity of the served area) that fulfills the requirements of decentralized production and consumption is therefore of a paramount importance~\cite{HVELPLUND20062293}. Flexibility, on the other hand, can be enabled by facilitating Demand Response~\cite{ALI2014}, which is an important element in the prosumer-centric energy market. 
While previous works have highlighted the need for demand response (DR) markets they have not considered small-scale prosumers as market participants in a local market. One example considers stakeholders such as DSO, TSO (as \emph{buyers}), aggregators and customers (as \emph{sellers})~\cite{tNguyen2011}. However, market interactions down to the level of prosumers are not considered. Another work presented an effort-based DR service where the DR participants are benefited against the time of their \emph{stalling effort} and showed that by doing so, the system achieves socially efficient allocations~\cite{CLA2016}. However, the implications of such an innovative design on the real electricity market scenarios are not entertained. 

Coordination of flexible resources/devices while aggregating their flexibility usually requires a detailed understanding of the underlying characteristics~\cite{ALI2014} of those devices which may be hard to obtain in many scenarios and thus hinders the scalability of the system. A considerable amount of research is focused on the management of heterogeneous devices (e.g. PV generator, storages, gas micro-turbine) through a centralized controller~\cite{Kanchev2018}, while hierarchical controls of microgrid through several levels of control architecture\cite{Guerrero2011,chakra2016} show promising applicability in this domain.
Additionally, in order to optimally manage and control flexibility in demand side, previous works deploy aggregators~\cite{Agentis2016,Ottesen:2016dc} that essentially combine the available flexibilities and trade them on the customers' behalf. 
However, this is not a straightforward process and may have negative impacts such as increasing the imbalance of Balancing Responsible Parties (BRPs) and leading to engaging in further coordination between the participants~\cite{FlexStudy}.  Moreover, the underlying loss emerging from the balancing activity incurred due to energy-only trading are socialized, which means prosumers are priced equally regardless of their marginal contributions to the overall loss. Hence, it is critically important to improve the fairness of the balancing cost attributions. Consequently, the need for a general and fair yet effective coordination and control scheme for flexible resources through local market design becomes crucial.

In light of these related works and the narrative aforementioned, this paper contributes to the state of the art by:
\begin{itemize}
\item Proposing a real-time exchange mechanism that promotes efficient allocations in ECs, coordinating heterogeneous and individually-controlled flexible resources.
\item Proposing a payment structure -- integrated with the proposed mechanism -- that incentivizes the prosumers to participate in the exchange mechanism, rewards them according to their contributions to system loss reduction, and by doing so, increases the social welfare of the residential community.
\end{itemize}

The rest of the paper is organized as follows: Section~\ref{sec_coord_machanism} describes the  prosumer marketplace and outlines the operation necessary for the exchange mechanism. The proposed control and exchanged mechanism is detailed in Section~\ref{sec_opt_ex_control}.
Section~\ref{sec_coi} elaborates the prosumers' utility and the fairness of the mechanism. A simulation case study is presented in Section~\ref{sec_case_study}. Finally, Section~\ref{sec_conclusion} concludes the paper with a glimpse of possible follow-up research.

\section{Prosumer Marketplace and Operation Outline}
\label{sec_coord_machanism}
In this section, we describe a real-time exchange mechanism that operates on a local \emph{prosumer marketplace} with associated market offers that are necessary for the market formation. Prosumers in a cooperative are represented by software agents, therefore, we will use the terms agent and prosumer interchangeably. While we discuss it in the context of storage flexibility from batteries, it is applicable to the flexibility of demand response as well.


In this section, we describe a real-time market with market offers that are necessary for residential ECs. The market while operating locally, facilitates trading energy and flexibility (of consumption, commonly known as \emph{Demand Response}) among the prosumers. The prosumers of the cooperative are represented by software agents. Therefore, we will use the terms agent and prosumer interchangeably.
The prosumers are assumed to be equipped with PV and Battery (as flexible resources). To this end, we formalize the market offers and associated commodities. 

\subsection{Market offers}
A \emph{market offer} $o = (c, p, f)$ is defined as a tuple comprising a \emph{commodity} $c \in \mathcal{C}$ with a price tag $p$ and a \emph{flag} stating whether the offer is a bid ($f=1$) or an ask ($f=0$). The offer can be further extended by adding a quantity $n$, differentiating between limit orders and fill-or-kill orders $l$, and an expiry \emph{timestamp} $t_{exp}$.

\subsection{Commodity classes}
\label{sub_sec_com_class}
A commodity, in general term, is a marketable item (typically, a good or a service) produced to satisfy demand or need~\cite{o2003economics}. In the context of the electric power market, a commodity comprises power time series, giving rise to several interesting classes as services.

\begin{itemize}
\item Energy: $\mathcal{C} = (\mathbf{q}, \mathbf{t})$, where $\mathbf{q}$ and $\mathbf{t}$ are equal-length vectors of time and quantity respectively, defining a (possibly continuous) time series with $\forall i: q_i \geq 0$, with the convention that the last quantity is zero. 
\item Demand Response: $\mathcal{C} = (\mathbf{q}, \mathbf{t})$, where $\mathbf{q}$ and $\mathbf{t}$ are vectors of time and quantity respectively, defining a (possibly continuous) time series, possibly imposing the zero-sum constraint $\sum_i q_i = 0$ to make this class of commodities complementary to energy 
This constraint may simplify discrimination from energy commodities but may be counter-productive as it avoids demand response losses to be included in the commodity. However, since losses cannot be standardized, demand response is rather amenable to negotiation than auction-based commodity trading.

\end{itemize}



In this article, however, we limit the contribution to devising exchange mechanisms of \emph{Energy} commodity while keeping the \emph{Demand Response} activated by indirectly controlling the local flexibility.
The market protocol proceeds in rounds where every agent submits a bid (or ask), allocations are computed by matching bids and asks, establishing the market equilibrium. In this paper, however, we have assumed all bids (and asks) are cleared by a local exchange market. The exchange market clears the offers in near real-time and provides the allocations back to the agents. The operation outline is detailed in the following section.


\subsection{Operation outline}
\label{sub_sec_outline}
\begin{figure*}[h]
\centering
\includegraphics[scale=0.5]{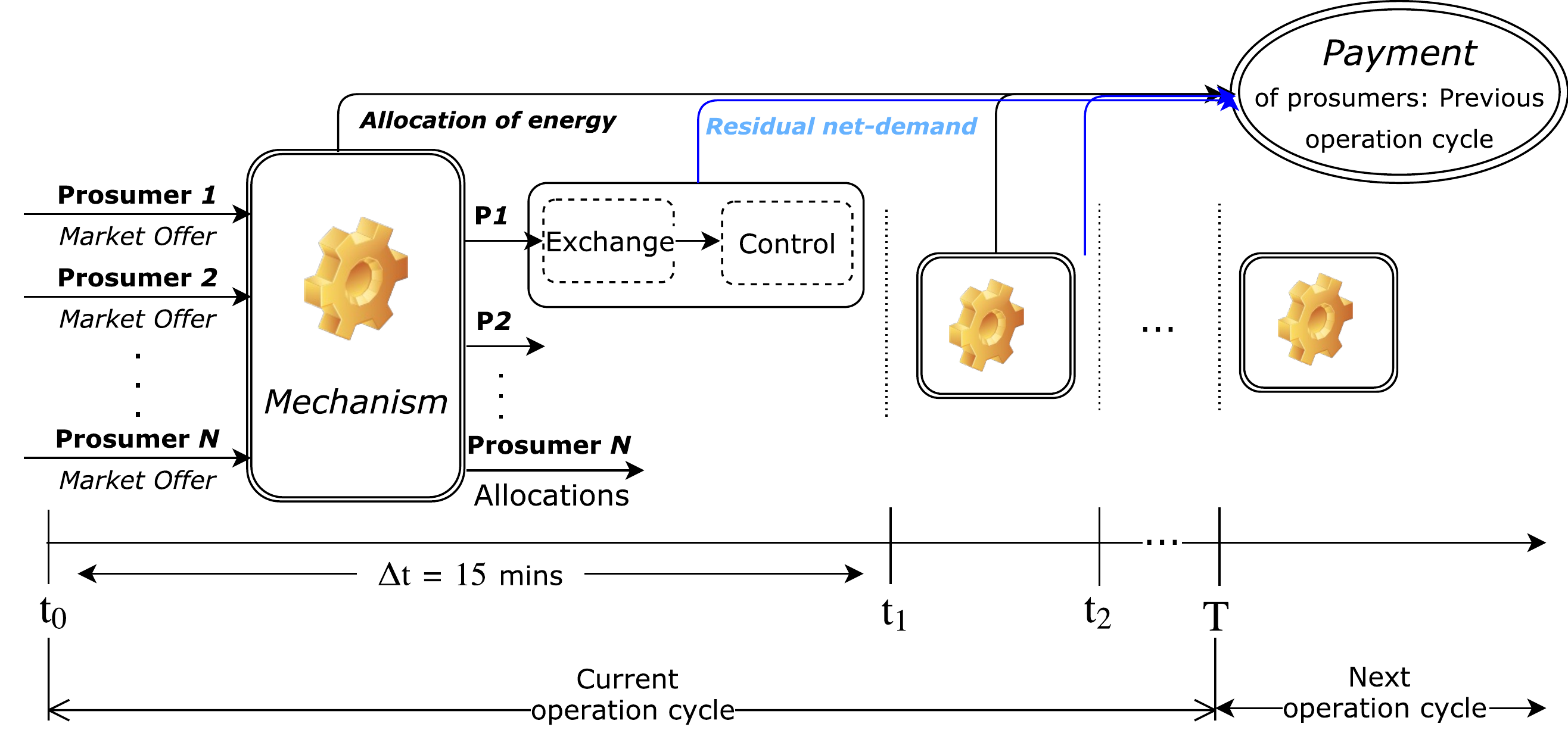}
\caption{High-level operational outline of the coordinating exchange mechanism: exchange and control (Section~\ref{sec_opt_ex_control}) followed by payment computation (Section~\ref{sec_coi}).}
\label{fig_operation_outline}
\end{figure*}

The high-level operational outline of the proposed mechanism is shown in Figure~\ref{fig_operation_outline}. In brief, the mechanism periodically accepts offers from prosumers, computes the allocated exchange quantities by matching offers, and responds back to the prosumers with those allocations. The additional payments are calculated retrospectively after a predefined \emph{operation cycle} of $T$ discrete time periods is completed. Specifically, the algorithm proceeds as follows:

\begin{itemize}
  \item \textit{Step 1:} For each time $t\in T$, repeat \textit{Step 1.1} to \textit{Step 1.5}.
  \begin{itemize}
    \item \textit{Step 1.1:} Collect the \emph{market offer}s from all prousmers, segregating \emph{bids} and \emph{asks} based on the embedded $f$ flag.
    \item \textit{Step 1.2:} Tabulate the quantities with round-trip efficiencies of each flexible devices of sellers (detailed in Section~\ref{sec_opt_ex_control}).
    \item \textit{Step 1.3:} Determine the \emph{allocations}, using the \emph{bids} and \emph{efficiencies} of the devices after applying the proposed exchange mechanism.
    \item \textit{Step 1.4:} Respond back to the prosumers with the allocations, where the prosumers realize the exchange and deploy batteries.
    \item \textit{Step 1.5:} Transact the prosumer-specific residual net energy with the Grid.
  \end{itemize}
  \item \textit{Step 2:} Determine the payments of prosumers as detailed in Section~\ref{sec_coi}. The form of payment includes the transactions with the Grid, realized energy exchange with peer prosumers, and allocated flexibility activation.
\end{itemize}

Next, we introduce the baseline cases of \emph{No flexibility} and \emph{Individual control}, and proceed to provide a brief description of the proposed exchange mechanism (i.e. \emph{Exchange and control}), which is elaborated in the subsequent section.

\begin{itemize}
\item \emph{No flexibility: } Considers the simplest case where there is no exchange and the flexibility is deactivated (i.e., batteries are not considered). The net-demand is always resolved by interacting with the main Grid.

\item \emph{Individual control: } The agents individually control their batteries by charging with excess PV and by discharging to serve their own load. Agents are not allowed to exchange energy. 

\item \emph{Exchange and control: } Agents are allowed to exchange energy. We assume a centralized exchange is responsible for computing allocations from participants' offers (see Section~\ref{sec_opt_ex_control}). An offer comprises the net demand and the \emph{round-trip efficiency} of their flexibility (e.g., battery efficiency). By doing so, our proposed mechanism is able to improve allocative efficiency by resolving flexibility needs with the most efficient means.
\end{itemize}

\section{Exchange and control mechanism}
\label{sec_opt_ex_control}

The optimal allocation for pooled resources is computed by solving a centralized Linear Programming problem. Let, $ex_{b}(s) \in \mathbb{R}^{+}$ be the energy traded from seller $s \in S$ to buyer $b\in B$; where $B$ and $S$ are the set of all buyers and all sellers (both including the Grid $\{\mathcal{G}\}$), respectively. Aligning with the \emph{commodity} as \emph{Energy} (defined in Section~\ref{sub_sec_com_class}), at any discrete time interval $t$, the buyers and sellers approach to the real-time exchange market by providing a quantity $q$ as $E_{b}(t)$ and $E_{s}(t)$, respectively. Therefore, for a buyer $b$ the bid structure looks like $c \leftarrow \left( q =E_{b}(t), t \right)$ and the \emph{market offer} is $o=\left( c, f=0\right)$.\footnote{The price component $p$ is omitted since the mechanism assumes clearing all bids and asks.} Similarly, for a seller $s$, the bid structure and the \emph{market offer} are represented by $c \leftarrow \left( q =E_{s}(t), t \right)$ and $o=\left( c, f=1\right)$, respectively. In addition to submitting the net demand to the centralized exchanger, the mechanism assumes that sellers submit the round-trip efficiency of their locally controlled flexible devices. Therefore, the bids are restructured as $c \leftarrow \left(q=<E_{s}(t), \eta_{s}>, t \right)$ where, $\eta_{s}$ be the round-trip efficiency of the battery (or other flexible resources) of seller $s\in S - {\mathcal{G}}$. The mechanism makes use of these submitted round-trip efficiencies to compute the allocation.
The objective is to maximize the local energy exchange and thereby minimizing the Grid exchange. 

In order to do so, for a particular seller $s$, the exchanger reformats $\eta_{s}$ to $\eta_{s}^{b},~\forall b \in B$ and assigns $\eta_{s}^{b} = \infty$ when $s:=\{\mathcal{G}\}$ or $b:=\{\mathcal{G}\}$. The transformation can be written as, $\forall s \in S$

\begin{align}
\eta_s^b = \left\{\begin{matrix}
\eta_s,  & \quad \text{if  } b\in B - \{\mathcal{G}\}\\ 
\infty,  & \text{otherwise.}
\end{matrix}\right.
\end{align}
After receiving the quantities as bids from prosumers, the exchange decides on the allocation that maximizes energy exchange locally (equivalently, minimizes energy transaction with the Grid\footnote{The mechanism does not actually allocate any Grid exchange with prosumers. Rather, it provides the optimal allocation among the prosumers.}). The following objective function serves the purpose of minimization of the Grid exchange with improved allocative efficiencies
\begin{align}
& \underset{ex}{\min} \sum_{b\in B}\sum_{s\in S} \eta_{s}^{b}\cdot ex_{b}(s).
\label{eq_eff_ex_con}
\end{align}
The above objective function is subjected to the following balancing equations for buyers and sellers, respectively.
\begin{align}
E_{b} = \sum_{s \in S} CM(b,s) \cdot ex_{b}(s) \quad \forall b \in B - \{\mathcal{G}\} \nonumber \\
E_{s} = \sum_{b \in B} CM(b,s) \cdot ex_{b}(s) \quad \forall s \in S - \{\mathcal{G}\},
\label{eq_ex_control_const}
\end{align}
where $CM$ is a ``boolean'' connectivity matrix representing possible trading restrictions due to the electrical network topology of the cooperatives. These constraints ensure that the demand (and the supply) is met entirely through the exchange with physically connected prosumers. 

The optimization problem (Eq.~\ref{eq_eff_ex_con}) with constraints (Eq.\ref{eq_ex_control_const}) essentially forms to an \emph{assignment problem}.
In this paper, we assume that the network is not restrictive, and it is therefore modeled by a fully-connected topology i.e. $CM(b,s)=1, \forall b \in B $ and $\forall s \in S$. Note that this mechanism is device-agnostic, i.e., it is not considering the (locally-controlled) storages' internal specification in computing allocations. 

The aforementioned optimization problem with associated constraints is solved by the Linear Programming technique.
Having said that, such a network topology is virtually possible by adapting advanced routing technology, e.g. Digital-Grid~\cite{ABE11} that provides every prosumer the platform to exchange power with every other prosumers in the cooperative facilitated by the installed Digital Grid Routers in suitable nodes. 

\section{Prosumers' Utility and Fairness of the Mechanism}
\label{sec_coi}
In this section, we describe the utility of prosumers by applying the coordination mechanism and go on showing how the mechanism increases the utility of prosumers, improving the \emph{social welfare}~\cite{moulin2004fair} and thereby resolving the \emph{conflict of interest} of the prosumers. For ease of description, we number the mechanisms as \emph{No flexibility} (0), \emph{Individual control} (1), and \emph{Exchange and control} (2), respectively.

Let $N$ be the number of prosumers and $T$ be the current \emph{operation cycle}. Further, let the demand and (PV) generation of agent $i$ at time $t$ be $d_{i}(t)$ and $pv_{i}(t)$, respectively. The net demand at period $t$ after applying mechanism $m$ and deploying the local battery of agent $i \in N$ is defined as:
\begin{align}
D_{i}^{m}(t) = d_{i}(t)-pv_{i}(t)-Ex_{i, -i}^{m}(t)+pb_{i}(t),
\label{eq_net_demand}
\end{align}
where $Ex_{i,-i}^{m}(t)$ is the total energy agent $i$ received from other agents $-i$ at time $t$ utilizing mechanism $m$. Therefore, $Ex_{i}$ essentially  is the agent specific decision allocated by the exchange described in Section~\ref{sec_opt_ex_control}. We suppose that surplus generated energy is not permitted (or at least not remunerated) to be fed into the grid, i.e. power flow from Grid $\mathcal{G}$ to agent $i$, $D_{i}^{m}(t)$ is positive, with excess energy spilled:
\begin{align}
D_{\mathcal{G}, i}^{m}(t) = [D_{i}^{m}(t)]^{+},
\end{align}
where $[x]^{+}=\textrm{max}\{x, 0\}$. Since \emph{mechanism 0} and \emph{mechanism 1} do not impose any exchange, $Ex_{i,-i}^{0,1}(t) = 0$ for all $i\in N$. The dispatched battery power of agent $i$ at $t$ is $pb_{i}(t)$, where $pb_{i}(t) < 0$ when the battery is discharged. When an agent switches from \emph{No flexibility} mechanism, the system encounters losses due to the round-trip efficiency of activated batteries. The overall loss contributed by $i$ after switching to \emph{mechanism} $m \in \{1, 2\}$ is given by
\begin{align}
L_{i}^{m} &= \sum_{t}^{T}\left ( D_{i}^{0}(t)-D_{i}^{m}(t) \right ) - \theta_{i}^{m}\nonumber \\ 
\quad & = \sum_{t}^{T}\left ( Ex_{i,-i}(t) - pb_{i}(t) \right )- \theta_{i}^{m},
\end{align}
where $\theta_{i}^{m}$ is an offset power dispatched from/to the battery to bring the final state of the charge to the initial one. Finally, total \emph{loss of the system} using mechanism $m$ is calculated as
\begin{align}
\mathcal{L}^{m} &= \sum_{i=1}^{N}L_{i}^{m} =-\sum_{i=1}^{N}\left ( \sum_{t}^{T}pb_{i}(t) + \theta_{i}^{m} \right ).
\label{eqn_loss}
\end{align}
In order to instill fairness in the mechanism, we adopt a measure, namely \emph{difference evaluation function}, which has been widely employed in multiagent learning~\cite{Colby2016}. In this context, the \emph{difference evaluation function} of an agent $i$ defines the marginal contribution of $i$ in reaching a joint system state (i.e. \emph{loss in the system}), $\mathcal{L}$ utilizing mechanism $m$. Formally,
\begin{align}
\mathcal{L}_{i}^{m} = \mathcal{L}^{m} - \mathcal{L}_{-i}^{m}.
\label{eqn_marginal_loss}
\end{align}
We define a \emph{component} function that rewards/penalizes an agent based on its marginal contribution to system loss (i.e. $\mathcal{L}_{i}^{m}$). It can be shown from Eq.~\ref{eqn_loss}, that the \emph{component} function links with the agent's individual battery activation and therefore, provides a way to measure the effect of battery activation and consequent profit sharing with other agents. Formally, for agent $i$ using mechanism $m=2,$\footnote{Mechanisms 0 and 1 do not engage in exchange and hence do not require \emph{component} function.} the \emph{component} function goes
\begin{align}
R_{i}^{m} = L_{i}^{m} - w_{i}^{m}\mathcal{L}^{m},
\label{eq_reward}
\end{align}
where $w_{i}^{m}$ constitutes the weights proportional to the contribution of $i$ using the following softmax function
\begin{align}
w_{i}^{m}=\frac{e^{\mathcal{L}_{i}^{m}}}{\sum_{j}e^{\mathcal{L}_{i}^{m}}},
\end{align}
where $\sum_{i}w_{i}^{m}=1$, which makes $\sum_{i}R_{i}^{m}=0$. It is an important criterion to fairness argument as the total rewards must be shared among all participated agents.

Finally, the utility of an agent $i$ by applying \emph{mechanism} $m$ is defined as the negative cost it incurs from the (total) payment to the Grid, payment from/to the exchange (with other agents), and the reward defined in Eq.~\ref{eq_reward}. For a constant price $p$, the utility function takes the following form
\begin{align}
u_{i}(m) = -p\left[ \sum_{t}^{T}D_{\mathcal{G}, i}^{m}(t) + \alpha\sum_{t}^{T}Ex_{i, -i}^{m}(t) - (1-\alpha)R_{i}^{m} \right],
\label{eq_utility}
\end{align}
where $\alpha=[0,1)$ represents the price component (as a fraction of $p$) an agent pays for exchanged energy. The complementary price component, i.e. $(1-\alpha)$) scales the \emph{Reward} and thereby controls the profit sharing between agents according to their flexibility activation. Inherently, the utilities of \emph{No flexibility} and \emph{Individual control} only contain the Grid payment component. Note, that the price $p$ is set out by the Grid and is in our setting identical for all prosumers in the EC for a particular \emph{operation cycle} $T$. The acceptability of the proposed mechanism depends on whether the proposed coordination mechanism ($m=2$) satisfies individual rationality of agents, i.e.
\begin{align}
u_i(m=2) \geq u_i(m=k),
\label{eq_u_max}
\end{align}
where $k \in \{0, 1\}$. The mechanism searches for the lowest $\alpha$ that satisfies the following constraint
\begin{align}
\sum_{t}\left( D_{\mathcal{G}, i}^{k}(t) - D_{\mathcal{G}, i}^{m}(t)\right)+R_{i}^{m}<\alpha\left(\sum_{t}Ex_{i, -i}^{m}(t)+R_{i}^{m}\right), 
\end{align}
for $m=2$, $k \in \{0, 1\}$ and $i=1,\dots, N$. Typical values of $\alpha$ stay within $0.75 \leq \alpha < 1.0$.
In order to validate the efficiency of the mechanisms in improving the social welfare, we define the following \emph{utilitarian social welfare} function
\begin{align}
sw_{m} = \sum_i^{N} u_i(m).
\label{eq_sw}
\end{align}
Moreover, we quantify the relative fairness of a mechanism based on the \emph{Nash social welfare} criterion, an established concept of fairness~\cite{moulin2004fair}, applied to the improvement above a baseline:
\begin{align}
f_{m|k} = \prod_i^{N} \left ( u_i(m) - u_i(k)\right ),
\label{eq_fairness}
\end{align}
where $m \neq k$. Utilizing Eq.~(\ref{eq_net_demand} -~\ref{eq_u_max}), it can be shown that $m=2$ improves both social welfare functions, i.e. Eq.~\ref{eq_sw} and Eq.~\ref{eq_fairness}, and hence balances efficiency and fairness. The next section supports this claim empirically.

\section{A Case Study of local energy exchange}
\label{sec_case_study}
In this section, we present experimental results of a case study applying the proposed mechanism. The experiments are conducted by taking samples of 10, 20 and 30 prosumers (so called ``scalability'') over 150 households from real residential data~\cite{PSData}. We present experiment results considering the \emph{loss reduction} and the \emph{payments and social welfare} after applying various mechanisms.
\begin{figure}[h]
\centering
\includegraphics[scale=0.40]{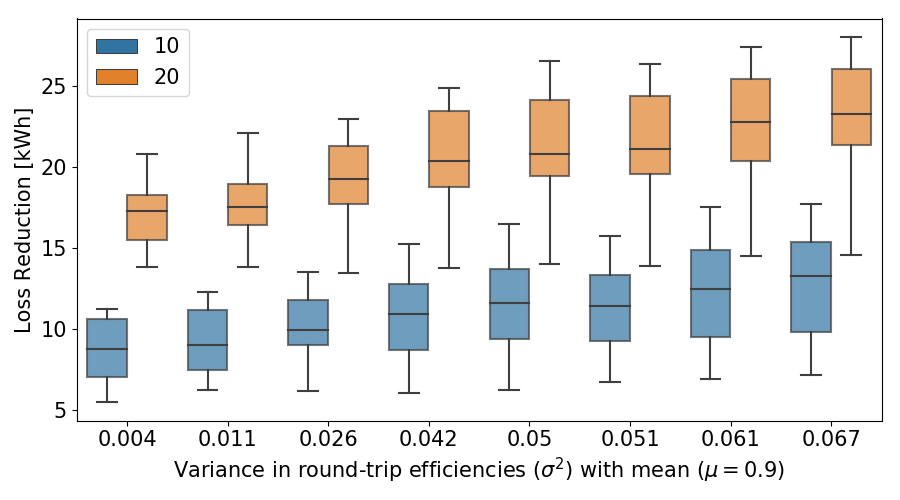}
\caption{System \emph{loss reduction} over \emph{Individual control} with increasing diversities in terms of number of prosumers (10, 20) and round-trip efficiencies.}
\label{fig_loss_vs_scale_static_eff}
\end{figure}
We use 10 and 20 prosumers cases for \emph{loss reduction} experiment, whereas 30 prosumers case is used for \emph{payment and social welfare} experiment. The prosumers are assumed to be equipped with a battery\footnote{A residential Lithium-ion battery of capacity 6.8kWh with rated discharging and charging power of 3kW and 1.3kW, respectively. The SOC is allowed to operate within 10\% to 90\% of the capacity. The degradation rate of the battery is fixed as 0.1\% of the capacity.}. In order to experimentally investigate the effect of ``diversity'' on system performance, the round-trip efficiencies of the flexible resources are varied by sampling from normal distributions.
The PV generation patterns of a set of randomly chosen prosumers are shifted uniformly with a span of [0,6] hours in order to form a local \emph{prosumer marketplace}, simulating diversified installation angles (e.g., east/west facing solar panels). The simulation period covers 7 days, during which the exchange is performed over each quarter-hourly interval.

\begin{figure}[h]
\centering
\includegraphics[scale=0.38]{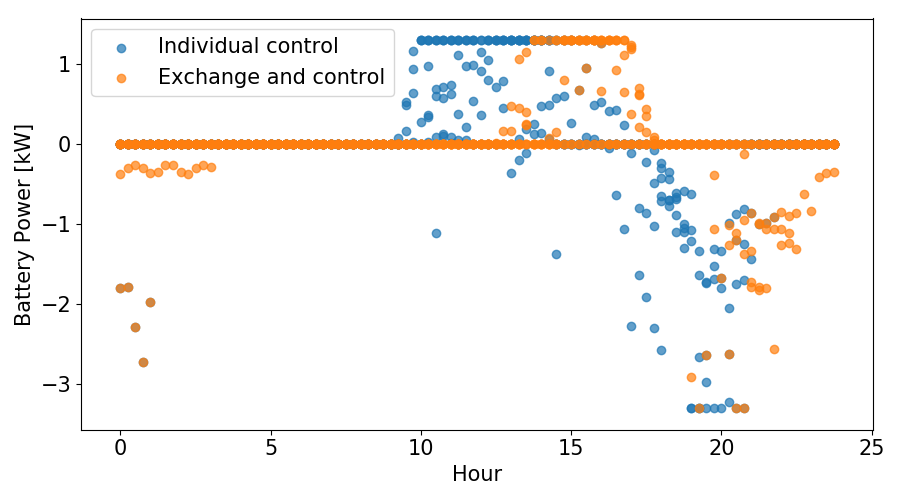}
\caption{Example battery power dispatch (of one prosumer) by applying \emph{Individual control} and \emph{Exchange and control} mechanism plotted over a day. The exchange mechanism yields a ``shifting'' in battery scheduling.}
\label{fig_pro_b_power}
\end{figure}

\subsection{Experiment: Loss Reduction}
\label{sec_experimental_loss}
Here, we analyze the performance of the proposed mechanism with respect to the \emph{loss reduction}, as described in Eq.~\ref{eqn_loss}. The experiments are performed over a number of trials, where each trial contains an independently chosen subset of the prosumers - the dataset of prosumers contains 150 prosumers. The improvement in loss reduction using the \emph{Exchange and control} mechanism over the \emph{Individual control} mechanism (i.e. $\mathcal{L}^{m=2}-\mathcal{L}^{m=1}$) for different number of prosumers with increased diversity -- measured by the variance in the distribution of round-trip efficiency of batteries -- is depicted in Figure~\ref{fig_loss_vs_scale_static_eff}.
The positive \emph{loss reduction} indicates that the \emph{Exchange and control} mechanism influences the flexibility activation indirectly and thereby results in a system-wise higher efficiency. As the scale of the EC grows, so does the dominance of \emph{Exchange and control} mechanism over the \emph{Individual control}. 
As for the experiment with ``diversity'', the \emph{Exchange and control} mechanism takes the round-trip efficiencies into consideration and reduces the loss further with greater diversity. An agent-specific residual battery power dispatch patterns, resulted from the \emph{Individual control} and the \emph{Exchange and control} mechanisms for 7 days (plotted in scatter on a 24-hour window) are shown in Figure~\ref{fig_pro_b_power}. The storage pattern, emerging from the \emph{Exchange and control} mechanism, shows a ``shifting'' tendency from that of the \emph{Individual control} mechanism, which is resulted from the agent-specific realized allocation. The experiments confirm that the proposed mechanism efficiently coordinates the local flexible resources while resulting in a higher \emph{loss reduction} when compared to \emph{Individual control}.

\begin{figure}[tb]
\centering
\includegraphics[scale=0.48]{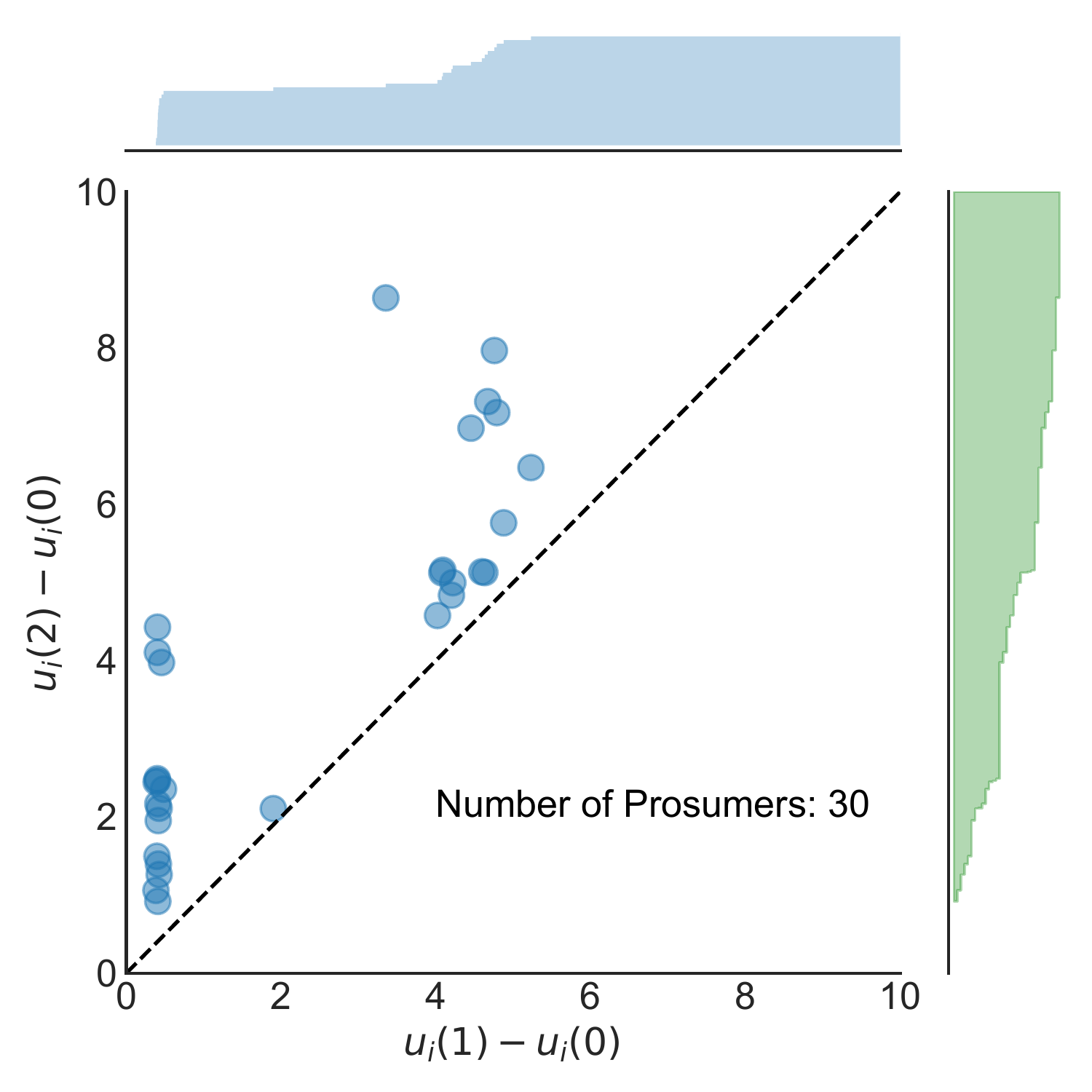}
\caption{Prosumer utility improvements of \emph{Individual control} (horizontal) and \emph{Exchange and control} (vertical) over \emph{No flexibility} baseline, with marginal \emph{cumulative} distributions. Prosumers are above the dashed equal improvements line, hence the new \emph{Exchange and control} mechanism dominates \emph{Individual control}, while also vastly improving relative fairness $f_{2|0} \approx 8.89 \cdot 10^{10} \gg f_{1|0} \approx 3.24 \cdot 10^3$.}
\label{fig_utility}
\end{figure}

\subsection{Experiment: Payment and social welfare}
\label{sec_conf_interest}
Lastly, we turn the analysis toward the payment strategy of the exchange mechanism, and how the strategy makes the mechanism preferable for all individual agents. Figure~\ref{fig_utility} presents the relative increase in utility (i.e. decrease in cost) for each prosumer of an EC with 30 prosumers, comparing the improvements of \emph{Individual control} and \emph{Exchange and control} over \emph{No flexibility}. As seen in the figure, $(u_{i}(2)-u_{i}(0))$ dominates over $(u_{i}(1)-u_{i}(0))$ by placing itself over the dashed line and by exhibiting a smoother \emph{cumulative} distribution. Therefore, it implies that the social welfare functions defined in Eq.~\ref{eq_sw} and Eq.~\ref{eq_fairness} are maximized by \emph{Exchange and control} mechanism. 
The \emph{Nash} social welfare function essentially balances the efficiency and the fairness. Therefore, it is evident that the \emph{Exchange and control} mechanism is efficient and fair with the proposed payment structure.


\section{Conclusion}
\label{sec_conclusion}
In this paper, our first contribution is a market mechanism -- operating on a near real-time \emph{prosumer marketplace} -- that ensures efficient allocation of flexible energy resources in residential Energy Cooperatives (ECs). Our \emph{Exchange and control} mechanism is device-agnostic as it does not require detailed modeling of the locally and independently controlled distributed flexible devices (e.g. batteries). 
The mechanism's payment function rewards participating prosumers based on their marginal contributions to the joint system states (i.e. \emph{loss reduction}), which encourages the prosumers to participate in the exchange market while also resolving the conflict of interest between prosumers. This results in an improved \emph{utilitarian} social welfare and improved fairness w.r.t. \emph{Nash} social welfare.
Due to the bidding and exchange execution in a near real-time fashion, the proposed mechanism inherently reduces the necessity of predicting demand and generation that are required for bidding in the day-ahead market.
A detailed case study over real demand/generation data for various scales of EC with infused diversifications from underlying flexible devices and a follow-up sensitivity analysis to achieve \emph{loss reduction} and improvement in social welfare has validated the efficiency and applicability of the proposed exchange mechanism in a residential EC setting.

\section*{Acknowledgment}
This research has received funding through the ERA-Net Smart
Grids Plus project Grid-Friends, with support from the European Union’s Horizon 2020
research and innovation programme, and is a part of Energy Transition Hub research at the University of Melbourne, Australia.



\end{document}